\documentclass[prl,amsmath,amssymb,floatfix,twocolumn]{revtex4}
\usepackage{amsfonts}
\usepackage{graphicx}% Include figure files

\begin{document}

\def\E{{\bf E}}

\def\J{{\bf J}}

\def\B{{\bf B}}

\def\Q{{\bf \Theta}}

\def\D{{\bf D}}

\def\r{{\bf r}}

\def\v{{\bf v}}

\def\dV{{\; \rm d^3}{\bf r}}

\def\curl{{{\; \rm curl}\; }}

\def\grad{{{ \rm grad}\; }}

\def\div{{{\rm div}\; }}

\def\p{{\bf p}}

\def\U{{\mathcal U}}

\def\dv{{\rm d^3}{\bf r}}

\def\Z{{\mathcal Z}}

\title{Local Molecular Dynamics with Coulombic Interactions}

\author{J\"org Rottler\footnote{current address: Princeton Institute for the Science and Technology of Materials, Princeton University, Princeton, NJ 08544} and A.C. Maggs}

\affiliation{Laboratoire de Physico-Chimie Th\'eorique, UMR CNRS-ESPCI
  7083, 10 rue Vauquelin, F-75231 Paris Cedex 05, France}

\pacs{02.70.Ns,87.15.Aa,41.20.Cv}

%\date{\today}

\begin{abstract} 
  We propose a local, ${\cal O}(N)$ molecular dynamics algorithm for
  the simulation of charged systems.  The long ranged Coulomb
  potential is generated by a propagating electric field that obeys
  modified Maxwell equations.  On coupling the electrodynamic
  equations to an external thermostat we show that the algorithm
  produces an effective Coulomb potential between particles. On
  annealing the electrodynamic degrees of freedom the field
  configuration converges to a solution of the Poisson equation much
  like the electronic degrees of freedom approach the ground state in
  ab-initio molecular dynamics. 
\end{abstract}

\maketitle

Coulomb's law for the interaction between two charged particles is
generally presented as a static limit of Maxwell's equations valid
after all transients have decayed to zero \cite{jackson}. Due to the
difference between the signal propagation speed, $c$, of
electromagnetic radiation and excitations in condensed matter, almost
all particle-based numerical simulations of materials employ the
approximation of static, instantaneous interactions $(c=\infty)$. This
approach has some disadvantages. Since the electrostatic potential is
the unique solution of Poisson's equation, even the slightest motion
of particles requires a {\em global} recalculation of the
electrostatic potential at every time-step; this calculation 
can dominate the computational effort \cite{schlick} and represents a
major bottleneck for the development of efficient multiprocessor
codes.

One might wonder whether more efficient code results from a
formulation that allows one to reduce the propagation speed, but still
maintains a sufficiently large separation of time scales in a manner
familiar from ab-initio molecular dynamics \cite{car}. The ratio of
the rms particle velocity $\bar v$ to $c$ would play the role of an
optimization parameter much like the ratio of electron to nuclear
masses in quantum chemistry.  In order to change $c$, one would
simulate the evolution of the coupled particle-electromagnetic system
as is routinely done in plasma physics \cite{buneman}.  Such a
treatment has the great advantage of only requiring {\em local}
operations, but requires an enormous reduction of c in order to be
efficient.  However, with such a dramatic reduction of $c$ the
electric field will not follow the particle motion adiabatically. In
this limit there is no guarantee that the correct thermodynamic
ensemble is generated.  A fundamental question thus arises: Is
Coulomb's law {\sl just} a static limit of the Maxwell equations or is
the law more general?

Recent work on Monte-Carlo algorithms \cite{acm,acm2,joerg} has shown
that the correct thermodynamic potential is found even if the
particles and fields propagate at the same rate:
If one writes that the energy of an electric field ${\E}$ is $\U =
\int \E^2/2 \;\dv$, then a Gibbs distribution characterized by
interactions in $1/r$ is generated from the constrained integral
\begin{equation}
 \Z(\{ \r_i\}) = \int {\cal D}\E\,\prod_\r 
\delta(\div  \E(\r) - \rho(\{ \r_i\})) 
e^{- \U/k_BT}.
\label{eq-Z}
\end{equation}
where the charge density, $\rho({\bf r})=\sum_ie_i\delta({\bf r}-{\bf
r}_i)$; the charge of the $i$'th particle is $e_i$. We work in
Heaviside-Lorentz electromagnetic units where
$\epsilon_0\equiv\mu_0\equiv 1$.  The constraint in the
$\delta$-functions is Gauss' law.  In the electrostatic limit,
$\E=-\grad \phi_p$, where $\phi_p$ is the solution of Poisson's
equation with source $\rho$, but in general $\E=-\grad \phi_p+\E_{\rm
tr}$, where $\E_{\rm tr}$ is an arbitrary transverse or rotational
vector field.  By changing variables and integrating over $\E_{\rm
tr}$, one immediately sees \cite{acm, acm2} that the longitudinal
field components result in a Coulombic partition function, while the
contribution from the transverse components merely multiply this
partition function by a constant. Thus Coulomb's law is valid even in
the presence of a fully equilibrated and dynamic transverse
electrostatic field.  This result is non-trivial since the retarded
interaction between two charges is not simply $1/r$ \cite{jackson}.

This Letter implements two molecular dynamics algorithms which sample
Eq.~(\ref{eq-Z}).  We show that our algorithms generate the correct
thermodynamic potential and discuss the consequences of lowering $c$
for time dependent correlations. The algorithms involve only {\sl
local operations} on the field degrees of freedom and the CPU time per
integration step scales linearly with the number of particles $N$.
The most obvious choice for sampling Eq.~(\ref{eq-Z}) is to directly
integrate Maxwell's equations together with Newton's equations for the
particles coupled to the electromagnetic field:
\begin{eqnarray}
\dot \B &=& -c \,\curl \E, \quad\quad\quad
%\dot  m \v_i = e_i \E(\r_i)/m_i 
m_i\dot{\bf v}_i = e_i {\bf E}({\bf r}_i)
\nonumber \\
\dot \E &=& c \,\curl \B - \J,\quad \quad
\dot \r_i =  \v_i  
\label{eq-motion}
\end{eqnarray}
$\J$ denotes the electric current due to the particle motion.  As in
the electrostatic limit, we have dropped the Lorentz force $e_i( \v
\times \B)/c$ in the equation for $\dot \v_i$
\cite{nonham-comm}.

Evolution of the system is described by a map in phase space. If the
Jacobian of this map is unity then there is a conserved density, just
like in Hamiltonian mechanics where Liouville's theorem can be
applied: Let us integrate Eqs.~(\ref{eq-motion}) through a time step
$\delta t$ and evaluate the Jacobian of the transformation
\begin{math}
  {\partial x'_i / \partial x_j}
\end{math}
where $x_i$ denotes any one of the variables in
Eqs.~(\ref{eq-motion}). Since we have $\partial x'_i/\partial x_i=1$,
{\sl i.e.} the diagonal elements of this matrix are unity we find that
the Jacobian is $1+O(\delta t)^2$. Thus in the limit of small time steps
the Jacobian is preserved. This implies that a measure conserved by
the dynamics is
\begin{equation}
d\mu= \prod_{i,\alpha} dr_{i,\alpha}\;dv_{i,\alpha} 
\prod_{\r,\beta} dE_{\r,\beta} \; dB_{\r,\beta} 
\end{equation}
where the products are over the particles and then all space.  We note
that generalized Liouville dynamics have turned out very useful
recently in the construction of new thermostating methods for particle
simulation \cite{tuckerman}.  Since the equations of motion conserve
the energy
\begin{math}
\U_m = \sum {m \v^2_i / 2} + \int \dv 
\left\{ 
{\B^2/ 2} + {\E^2/ 2}
\right \}
\end{math}
we deduce that the partition function for this system is
\begin{math}
  \Z = \int d\mu\; \delta(U_m-U_0) \times \delta({\rm constraints})
\end{math}
where the $\delta$-function includes all the constraints and
conservation law inherent in the two Maxwell equations of
Eqs.~(\ref{eq-motion}).

A standard hypothesis of ergodicity would lead us to guess that we now
sample Eq.~(\ref{eq-Z}). However, the full Maxwell equations have
associated with them many independent conservation laws in addition to
the Gauss condition \cite{invariants}, most importantly $\div \B=0$.
If we simply integrate the Maxwell equations we have to supplement the
Gauss constraint in Eq.~(\ref{eq-Z}) with many other constraints in
such a way that the analytic formulation of the partition function
becomes intractable.  We should reduce the symmetries and conservation
laws inherent in Maxwell's equations, leaving just Gauss' law; we do
this by transforming to a constant temperature ensemble and coupling
the electromagnetic field to thermostats to improve the ergodicity of
the field degrees of freedom.

We modify two of the equations of motion to
\begin{eqnarray}
\dot m \v_i = q_i \E(\r_i)-\gamma_1 \v_i +\vec \xi_1, \nonumber \\
\dot \B = -c\, \curl \E -\gamma_2 \B + \vec \xi_2, 
\label{eq-motion2}
\end{eqnarray}
where the damping $\gamma_j$ and the noise $\vec\xi_j$ are related by
the fluctuation dissipation theorem.  The equation for the particle
velocity is entirely conventional in molecular dynamics, that for the
magnetic field less so.  The noise $\vec \xi_2$ on the magnetic field
degrees of freedom is completely general; it does not satisfy $\div
\vec \xi_2=0$.  Due to the coupling of $\B$ to the random noise it is
ergodic, as is $\v_i$.  Introduction of the noise has destroyed the
unwanted constraints arising from Maxwell's equations.  However taking
the divergence of
\begin{math}
  \dot \E = c\,\curl \B -\J
\end{math}
we see that Gauss' law is still valid for the thermalized equations of
motion if the equation of continuity $\div \J + \dot \rho=0$ holds,
and we start the simulation with an initial condition consistent with
Gauss' law.

One can check that the distribution
\begin{math}
  P_0=e^{-\beta U_m}
\end{math}
is a fixed point of the thermalized equations. Such a demonstration is
often taken as being a sufficient criterion for a thermostat in
physics; due to the presence of the conservation laws in Maxwell's
equations we have examined convergence of the dynamics by studying the
function \cite{young} $H= \int {(P(t)-P_0)^2 / P_0} \;d\mu$ and
calculating the dynamics of $H$ with a Fokker-Planck equation for the
full distribution function $P$.  By requiring that $\dot H=0$ we find
that $P$ must converge to a function of the form
\begin{math}
  P= {\mathcal A}(\E, \r_i) P_0
\end{math}
where
\begin{equation}
\left( \int \dv\; \left(c\,\curl \B -\J \right)\cdot  {\partial \over \partial \E}  
+ \sum_i  \v_i  \cdot {\partial \over \partial \r_i}  \right ) {\mathcal A}
=0 
\label{eq-A} 
\end{equation}
The condition that this equation is valid for arbitrary $\B$ leads to
the conclusion that ${\mathcal A}$ is a functional of only $\div \E$.
The general solution of Eq.~(\ref{eq-A}), is then a general functional
of $\div \E - \rho$:
\begin{math}
  {\mathcal A} = \bar A[\div \E(\r)-\rho(\{\r,\r_i\})].
\end{math}
Choosing $\div \E -\rho=0$ as a conserved initial condition then leads
to the required result $ P(t) \rightarrow P_0$. With the weight $P_0$
and the measure $d\mu$ we integrate over the Gaussian variables $\B$
and $\v_i$ and reproduce the required partition function
Eq.(\ref{eq-Z}) independent of $c$.

In our implementation of the algorithm, particles of mass $m$ move in
the continuum.  We interpolated the charges onto the $L^3/a^3$ nodes
of a cubic mesh using 3rd order B-splines, distributing the charge of
each particle onto 27 nodes \cite{joerg}; higher or lower order
schemes are possible if they conserve charge.  The same mesh is used
to discretize the field equations, and the electric field $\E$ is
associated with the $3L^3/a^3$ links. Groups of 4 links are grouped
into plaquettes, and the magnetic field $\B$ lives on the $3L^3/a^3$
plaquettes \cite{yee}. The particles interact in addition with a
shifted Lennard-Jones potential of scale $a$ truncated at its minimum,
$r_c= 2^{1/6} a$. This discretization is equivalent to a standard
7-point discretization of the Laplacian \cite{poisson}.

In order to integrate the equations of motion (\ref{eq-motion}) of the
coupled field/particle system, we use a velocity Verlet scheme. First
we advance the magnetic field $\B$ together with the particle
velocities to midpoint. Then, the values of the $\B$-field and the
velocities are used to advance the electric field $\E$ and the
positions $\r_i$.  Finally, the Langevin thermostats
Eqs.~(\ref{eq-motion2}) are applied and the $\B$-field/velocity moves
completed.  We have used a standard time-step $\delta t=0.01\tau$,
where $\tau=\sqrt{ma^2/\epsilon}$ is the unit of time and
$\epsilon=e^2/a$ the unit of energy.  The same time step is used for
both the particle and field equations. The Courant criterion for the
stable integration of Maxwell's equations is $\delta
t<a/\sqrt{3}c$. For the values of $c$ used in this Letter this
criterion is always satisfied and never limits $\delta t$.

The success of the method relies on implementing constraint conserving
couplings between particles and fields as well as exact local charge
continuity.  Motion of a particle leads to a local (finite)
fluctuation of the interpolated charge density, $\Delta \rho_l$.  From
this we construct a {\sl local current} $\J_l$ such that $\div \J_l =
-\Delta \rho_l$.  We decompose the displacement of a particle $\Delta
\r_i$ into a (time reversible) sequence of steps $\{\Delta x/2, \Delta
y/2, \Delta z, \Delta y/2, \Delta x/2\}$ \cite{leimkuhler}.  Each
substep in a direction $\alpha$ leads to a current in only those links
parallel to $\alpha$ \cite{current-comm}. The field update is then
slaved to the current $\Delta \E = -\Delta \J_l$. The force acting on
the particles is found from the principle of virtual work: A
fluctuation of a particle position $\delta \alpha$ induces a local
charge fluctuation and thus a {\sl local} current.  The force is just
$f_\alpha= - \delta U_m/\delta \alpha$ \cite{force-comm}. This
prescription for the force leads to the usual electric force $f_{\rm
el} = e_i \langle \E(\r_i)\rangle$, where $\langle \E(\r_i)\rangle$ is
a local average of the electric field which depends on the exact form
of the interpolation of the charge density to the lattice; this is the
origin of the acceleration in Eq.~(\ref{eq-motion}).

\begin{figure}[tb]
\includegraphics[width=8cm]{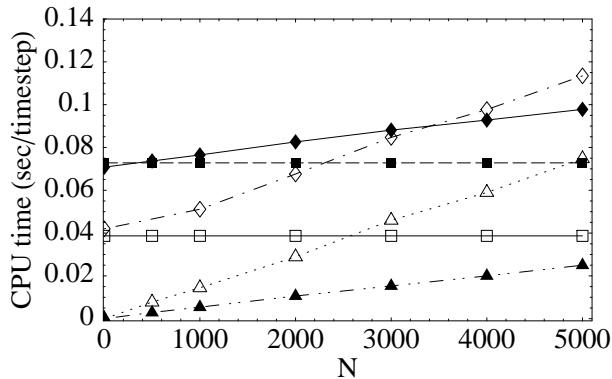}
\caption{\label{fig-linear} Time for the field integration
$(\square)$, field-particle couplings $(\triangle)$ and total time
$(\lozenge)$ in a system of size $L=30a$ as a function of $N$ on an
single AMD Athlon CPU. Also shown are results for the same system
treated with the P3M method using a charge interpolation of the same
order (filled symbols, $\blacksquare$ corresponds to the total
reciprocal work). }
\vspace*{-0.3cm}
\end{figure}

The use of a lattice to discretize the Maxwell equations leads to
artefacts in the interaction and self energy of the particles. These
artefacts are removed using dynamic subtraction \cite{joerg}.  A
scalar field that couples to the interpolated charges via the energy
functional
\begin{math} {\cal F}_Y [\psi]=
  \int \left[ {1 \over 2}[ (\nabla \psi)^2 + \mu^2\psi^2] - \rho \psi
  \right] d^3 r\label{yuk-eq}
\end{math}
leads to an effective interaction between particles of the form
\begin{math}
\label{yukpot-eq}
V_Y(\r) = - {e_i e_j e^{-\mu r} / r},
\end{math}
which when added to the direct Coulomb interaction regularizes the
short distance singularity in $1/r$. In our molecular dynamics code
this field obeys the equation of motion
\begin{equation}
\label{yuk-thermo}
{1\over c^2} {\partial^2 \psi \over \partial t^2} = \nabla^2 \psi -\mu^2 \psi +\rho- 
\gamma_3  {\partial \psi \over \partial t} + \vec\xi_3
\end{equation}
The Yukawa force $f_Y=+e_i\langle\grad \psi\rangle$ on the particle
comes form a local average consistent with the virtual work
principle, and the total force reads $f=f_{\rm el}+f_{\rm Y}+f_{\rm
LJ}$.  We correct for the Yukawa potential by adding an extra analytic
Yukawa potential (with opposite sign) to the truncated Lennard-Jones
potential at short distances.

Our algorithm runs in two modes. In mode I, all auxiliary fields are
kept at the same temperature as the particles, and we generate the
correct thermodynamic interaction independent of $\bar v/c$.  We also
run in ``dynamical simulated annealing mode'' \cite{car} (mode II),
where we anneal the electric degrees of freedom on ${\bf B}$ to zero
temperature by removing the noise and using larger values for
$\gamma_2$ and $c$, while maintaining the finite temperature
thermostat on the particles. In this case, our code is closer to
traditional methods, because the field configuration converges to the
solution of Poisson's equation when particle motion stops. This mode
is clearly similar in approach to dynamic methods used in quantum
simulations \cite{car} with $\bar v/c$ a freely variable optimization
parameter.
\begin{figure}[tb]
  \includegraphics[width=8cm]{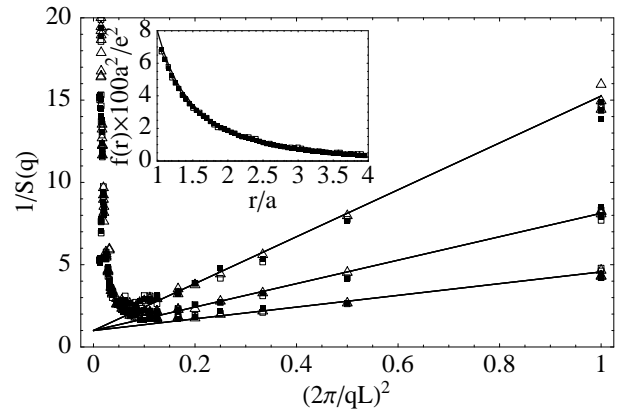}
\caption{\label{fig-struct}Static structure factor $S(q)= \langle
  s({\bf q})s(-{\bf q})\rangle,\,\, s({\bf
  q})=\frac{1}{\sqrt{N}}\sum_i e_i\exp{(i{\bf r}_i{\bf q})} $ of a
  charged symmetric electrolyte $(L=15 a)$; three densities $n=0.05
  a^{-3}$, $n=0.1 a^{-3}$, $n=0.2 a^{-3}$ and $T=\epsilon/4\pi k_B$
  using the present algorithm in mode I $(\square )$, $c=1a/\tau$;
  transverse electric field and particles thermostated to
  $T=\epsilon/4\pi k_B$, $\bar v/c\approx 0.3$ using
  Eqs.~(\ref{eq-motion2},\ref{yuk-thermo}). For $\blacksquare$, the
  electric field is damped, mode II, $c=32a/\tau$, $\bar v/c\approx
  0.01$. Also shown is a corresponding Monte-Carlo simulation
  $(\triangle)$. Solid lines: Debye theory. Inset: average
  instantaneous force for a pair of charges, $L=8a$ for modes I and
  II. Solid curve: $f=-dV/dr$, where $V=- 1/4\pi r-
  r^2/6\epsilon_0L^3$ from Ewald summation \cite{fraser}.  Lattice
  artefacts removed using dynamic subtraction, $\mu=0.9/a$. }
\vspace*{-0.3cm}
\end{figure}

Our numerical tests of the algorithm begin with a direct verification
of the $\cal{O}(N)$ scaling in Fig.~\ref{fig-linear} (open
symbols). While the effort to integrate the auxiliary fields only
depends on the number of grid points, the work required for the
particle-field couplings rises linearly with $N$. At the ``working
density'' of a typical biomolecular simulation using a 0.1 nm grid and
particle volumes of $\mathcal{O}(0.01 {\rm nm}^3)$ \cite{poisson}
(corresponding to $N\sim 3000$ in Fig.~\ref{fig-linear}), both parts
of the simulation contribute equally to the total time. Note in
particular that the field integration is twice as fast as the
reciprocal part of a standard Fourier-based P3M-implementation
\cite{plimpton2} (filled symbols). The expense for the particle-field
couplings is higher in our code, but optimizations in the prefactor
can be expected.

Next, we performed checks on the correctness of the method by placing
two particles of opposite charges in a box and measured the
instantaneous force.  We compare the results to the short range
expansion of the analytic Ewald summation in the inset of
Fig.~\ref{fig-struct} and find excellent agreement for both modes.  We
also simulated a globally neutral electrolyte composed of positive and
negative particles in order to observe the phenomenon of Debye
screening. At high enough temperatures, we expect a static
charge-charge structure factor of the form $S(q)=
{e^2q^2}/(\kappa^2+q^2)$ where the inverse Debye screening length
$\kappa^{2}=ne^2/k_BT$, $n$ is the density of charge carriers.
Fig.~\ref{fig-struct} compares $S(q)$ for several densities with this
expression and also to results measured with our related Monte-Carlo
algorithm \cite{joerg}.  Again we find good agreement. As described in
\cite{joerg} we also have studied the dispersion law of charge and
density fluctuations in a homogeneous electrolyte and were able to
confirm the fast equilibration of density-density and charge-charge
correlations.
\begin{figure}[tb]
  \includegraphics[width=8cm]{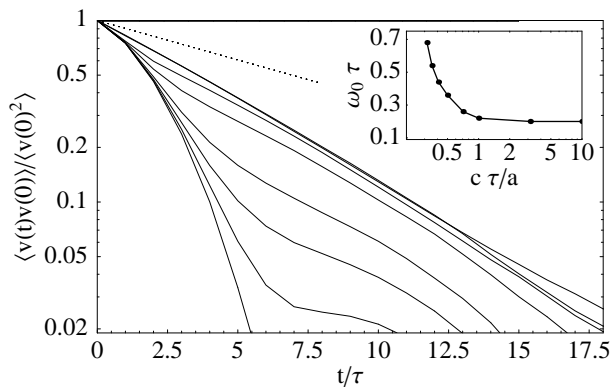}
\caption{\label{fig-autocorr} Velocity autocorrelation functions for
  $L=10a$, 50 particles, mode I. Intrinsic damping
  $\gamma_1=0.1\tau^{-1}$, dotted line: $\exp{(-0.1t/\tau)}$. Solid
  lines from left to right: correlations for $c=0.32, 0.35, 0.41, 0.5, 0.7, 1, 3.2, 10
  a/\tau$. Inset: initial decay rate $\omega_0$ from exponential fits.}
\vspace*{-0.3cm}
\end{figure}

In Fig.~\ref{fig-struct}, we showed that the algorithm generates a pair
potential equivalent to that of the Ewald summation for two very
different values of $\bar v/c$.  However, particle dynamics are
clearly sensitive to $c$. As a simple illustration, we show
velocity-autocorrelations $\langle {\bf v}_i(t){\bf v}_i(0)\rangle $
for several values of $c$ in Fig.~\ref{fig-autocorr}. For very dilute
systems the autocorrelation is dominated by the damping constant
$\gamma_1$ in the Langevin thermostat, and $\langle {\bf v}(t){\bf
v}(0)\rangle \sim \exp{(-\gamma_1t)}$. At higher densities, this decay
is modified by collisions, and $\langle {\bf v}(t){\bf v}(0)\rangle$
has a faster decay.  The curves in Fig.~\ref{fig-autocorr} are in this
density regime and are sensitive to the intrinsic dynamics of the
particles.  For $c<1 a/\tau$, correlations are modified by the low
propagation velocities, but saturate to a common curve for larger
values of $c$.

We have simulated a charged system interacting via Coulomb forces
using an algorithm in which the speed of light is a free variable. If
$\bar v/c$ is large the dynamics generate the correct statistical
mechanical ensemble for Coulomb interacting particles, but dynamic
correlations are modified. On increasing $c$ both the statistical
mechanical and the local dynamical properties are reproduced
correctly. The structure of our code is very similar to ``particle in
cell'' plasma codes which are rather easy to implement on large
multiprocessor computers with limited interprocessor bandwidth. We
therefore expect that on many processors, our algorithm can be
competitive with other fast electrostatic methods including Fourier
\cite{schlick}, multigrid \cite{darden2}, and fast multipole methods
\cite{greengard}.  Our method also generalizes naturally to situations
with spatial dielectric inhomogeneities, which cannot be solved using
Fourier techniques and non-standard boundary conditions, e.g.
irregulary shaped volumes.

We thank Burkhard D\"unweg for advice and encouragement in the
implementation of this method.

%\bibliography{mc}

\begin{thebibliography}{20}
\bibitem[{non()}]{nonham-comm}
\bibinfo{note}{This considerably simplifies the implementation of our code, but
  makes the equations non-Hamiltonian.}
\bibitem[{cur()}]{current-comm}
\bibinfo{note}{Explicitly, the electric field on the link departing from site
  $j$ in the (positive) $\alpha$-direction changes by $\Delta E_j^\alpha=\Delta
  E_{j-1}^\alpha +\Delta q_j^\alpha/a^2$, where $\Delta E_{j-1}^\alpha$ is the
  change in the field entering site $j$ and $\Delta q_j^\alpha$ the change of
  the partial charge on site $j$ due to particle motion in the
  $\alpha$-direction.}
\bibitem[{for()}]{force-comm}
\bibinfo{note}{Explicitly, the force acting on particle $i$ is
  $F^\alpha=-e_i\sum_{j}E_j^\alpha\partial E_j^\alpha/\partial r_{\alpha}$,
  where the sum runs over the interpolating sites before and after the particle
  motion.}
\end{thebibliography}

\end{document}